\documentclass[10pt,letterpaper]{article}
\usepackage[top=0.85in,left=2.75in,footskip=0.75in]{geometry}

\usepackage{amsmath,amssymb}

\usepackage{changepage}

\usepackage{textcomp,marvosym}

\usepackage{cite}

\usepackage{nameref,hyperref}

\usepackage[nopatch=eqnum]{microtype}
\DisableLigatures[f]{encoding = *, family = * }

\usepackage[table]{xcolor}

\usepackage{array}

\newcolumntype{+}{!{\vrule width 2pt}}

\newlength\savedwidth

\raggedright
\usepackage[aboveskip=1pt,labelfont=bf,labelsep=period,justification=raggedright,singlelinecheck=off]{caption}

\makeatletter
\renewcommand{\@biblabel}[1]{\quad#1.}
\makeatother

\usepackage{lastpage,fancyhdr,graphicx}
\usepackage{epstopdf}
\fancyheadoffset[L]{2.25in}
\fancyfootoffset[L]{2.25in}
\newcommand{\starsim}{\texttt{Starsim} }
\newcommand{\covasim}{\texttt{Covasim} }
\newcommand{\fpsim}{\texttt{FPsim} }
\newcommand{\hpvsim}{\texttt{HPVsim} }
\newcommand{\er}{Erd{\H o}s-R{\'e}nyi }
\newcommand{\ba}{Barab\'asi--Albert }

\begin{document}
\vspace*{0.2in}

\begin{flushleft}
{\Large
\textbf\newline{Noise-free comparison of stochastic agent-based simulations using common random numbers} 
}
\newline
\\
Daniel J.~Klein\textsuperscript{1*},
Romesh G.~Abeysuriya\textsuperscript{2},
Robyn M.~Stuart\textsuperscript{1\ddag},
Cliff C.~Kerr\textsuperscript{1}
\\
\bigskip
\textbf{1} Institute for Disease Modeling, Bill \& Melinda Gates Foundation, Seattle, WA, USA
\\
\textbf{2} Burnet Institute, Melbourne, Victoria, Australia
\\
\bigskip

%
%

\ddag Contractor on assignment




* Daniel.Klein@gatesfoundation.org

\end{flushleft}
\section*{Abstract}
Random numbers are at the heart of every agent-based model (ABM) of health and disease. By representing each individual in a synthetic population, agent-based models enable detailed analysis of intervention impact and parameter sensitivity. Yet agent-based modeling has a fundamental signal-to-noise problem, in which small changes between simulations cannot be reliably differentiated from stochastic noise resulting from misaligned random number realizations. We introduce a novel methodology that eliminates noise due to misaligned random numbers, a first for agent-based modeling. Our approach enables meaningful individual-level analysis between ABM scenarios because all differences are driven by mechanistic effects rather than random number noise. We demonstrate the benefits of our approach on three disparate examples. Results consistently show reductions in the number of simulations required to achieve a given standard error with levels exceeding 10-fold for some applications.

\section*{Author summary}

We present new computational methodology that addresses a longstanding signal-to-noise problem in agent-based modeling that arises when comparing simulation outcomes. With the traditional approach that we and other modelers have used for decades, random draw misalignment between simulations results in high variance and implausible differences, complicating impact evaluation, parametric sensitivity, and scenario analysis. Our new method achieve perfect alignment of random draws between simulations, thereby preventing stochastic branching entirely. Similar ideas have been demonstrated for simple cohort models, but those techniques did not work for key aspects we need in disease modeling like dynamic populations, births and in-migration specifically, and agent-to-agent interactions, as needed for pathogen transmission. We tested our new methodology on three use cases and found it has many benefits including dramatic reductions in the number of simulation replicates required for some applications. We believe that practitioners both within and beyond the field of computational epidemiology will benefit considerably from this improved approach to agent-based modeling.


\section*{Introduction}\label{s:intro}

Within the field of computational epidemiology, computer models are used to guide decision making by predicting the future course of disease burden, assessing data gaps and the value of new information, and quantifying the potential impact of a diverse suite of possible interventions. The structure and level of detail represented within a disease model should be fit for purpose based on the motivating questions and available data. To this end, numerous modeling paradigms have been developed and leveraged ranging from deterministic compartmental models to complex agent-based models (ABMs), which are inherently stochastic.

Simulation-based analysis quantifying the impact of interventions or the sensitivity of key outcomes to input parameters comes from evaluating the difference between simulations across two or more scenarios. These differences are straightforward to calculate for deterministic models, but significant challenges arise when evaluating differences between outputs of stochastic models, including agent-based models specifically.

The fundamental problem is that the difference between two simulations is composed of real (mechanistic) effects stemming from the change in model configuration plus stochastic noise. Configuration changes that result in small but meaningful differences in outcomes can be very challenging to quantify as the stochastic random number noise dominates the signal. Additionally, purely beneficial interventions and directional parameter shifts, like introducing a vaccine, can appear to \textit{increase} disease burden, challenging scientific communication. While such increases are certainly possible due to chaos-like mechanisms, traditional agent-based models over-estimate the frequency of such outcomes due to random number noise.

Several variance reduction techniques have been proposed in the literature to address this fundamental signal-to-noise problem in agent-based modeling and Monte Carlo simulation more generally~\cite{botev_variance_2017}. To understand these approaches, consider a disease model configured with two different inputs yielding outcomes $X$ and $Y$. The variance of the difference, $Z$, can be expressed as
\begin{align}
    \text{var}[Z] &= \text{var}[X] + \text{var}[Y] - 2\text{cov}(X,Y).
\label{eq:var}
\end{align}
To reduce the variance in the difference, it is possible to induce positive correlation between $X$ and $Y$ through the use of common random number seeds, a classic approach in simulation methodology~\cite{kahn_methods_1953}. In practice, however, the magnitude of the covariance term tends to be small relative to the variance terms despite the common random number seeds. While $X$ and $Y$ may be identical initially, the first difference causes a significant loss of correlation due to stochastic random number noise.

for example before an intervention takes effect in the counterfactual, the outcomes quickly lose covariance following the first difference due to stochastic random number noise.

The noise observed in differences is a result of a model design flaw that is challenging to overcome. Specifically, most agent-based simulations in epidemiology use a single centralized pseudo-random number generator (PRNG) for all stochastic realizations. A PRNG outputs a stream of random numbers that is deterministic and reproducible given the seed. However, even if two simulations with slightly different inputs use the same random number seed, as soon as one simulation uses a random number that the other does not, \textit{all} subsequent stochastic realizations could be different. The sequence of random numbers is the same, but the realizations are going to different purposes within the model. These differences manifest as noise when computing results. As a consequence, the impact of interventions and parameter changes can only be evaluated at the population level, even though the model representation is at the individual agent level.

Overcoming this fundamental limitation for general purpose agent-based simulation modeling has proven to be challenging; however, one promising approach is to use common random numbers (CRN)~\cite{kleijnen1974statistical}.  In a CRN-based approach, random numbers are still random, but the draw for each decision, by each agent, at each time is perfectly matched between simulations. In theory, the random number alignment in CRN eliminates stochastic noise from the difference of simulations, leaving only real (mechanistic) effects.

CRN techniques date back to Monte Carlo simulation in the 1950s, often applied to simple systems for which reusing a common random number seed was sufficient to achieve ``full'' common random number coherence~\cite{heikes_using_1976, conway_tactical_1963}. More recent applications of CRN include modeling of breast cancer~\cite{stout_keeping_2008}, health care systems and policy analysis~\cite{murphy_using_2013, cornejo_creating_2014}, and cost-effectiveness modeling of diarrheal disease control~\cite{flaxman_untangling_2017}.

While modeling applications using CRN consistently demonstrate benefits, no literature we could find solves CRN in general for agent-based simulation. The examples above suffer from two major limitations. First, the populations are closed in the sense that new agents cannot be born or otherwise added to the simulation. Second, agents are not able to interact with each other.

Common random numbers enable counterfactuals to be matched perfectly to baseline simulations. Perfectly matched counterfactuals have been demonstrated for compartmental models in epidemiology~\cite{kaminsky_perfect_2019}, but the methods do not apply to agent-based models.

We introduce new methodology for general purpose agent-based disease modeling that completely eliminates unwanted stochastic noise due to misaligned random numbers from the difference between simulations, thereby dramatically revealing the real signal. Our approach is based on common random numbers, and results in meaningful reductions in the variance of the difference between simulations. To the best of our knowledge, this is the first time that perfectly matched counterfactuals have been achieved in general-purpose agent-based modeling.

\section*{Materials and methods}\label{s:methods}


We achieve common random number alignment in an agent-based multi-pathogen co-transmission framework called \starsim using a number of innovations. This framework follows design patterns from specific disease-vertical models we have developed previously, including \covasim \cite{kerr2021covasim}, \fpsim \cite{fpsim}, and \hpvsim \cite{hpvsim}, but is intended to enable rapid composition of one or more health and/or disease modules and transmission networks. Importantly, \starsim is a fixed time step agent-based simulation framework that conceptually represents the agent population as a matrix. The matrix is composed of one row for each agent and has columns representing properties like age, sex, unique identifier (UID), pathogen-specific infection status, and much more. The matrix initially has $N$ rows, but will change dynamically over time as agents are born and die. The \starsim framework is written in Python and available as open-source software~\cite{starsim}. Here we describe each component of our approach. While a full implementation is available in \starsim, the approach can be adapted to any agent-based model.

\subsection*{Separate pseudo-random number streams for each decision}

Within agent-based modeling, a decision is any step that requires a random number to be drawn from a distribution. Typically these decisions address questions like: Does the agent get infected? How long is the incubation period? Will the infection be severe? Does the individual receive a vaccine on this time step? Etc. The results of these decisions govern the evolution of the simulation.

We assign independent PRNG streams to each and every decision, but note that each stream can be used to sample values for many/all agents. These decision-specific PRNGs have a unique name string that is hashed to create an integer offset to the single user-supplied random number seed. Thus, decisions that are shared between two simulations will receive the same seed, provided the overall random number seed is shared.

Within \starsim, each independent PRNG stream is implemented as a NumPy random generator of type \texttt{PCG64DXSM}~\cite{pcg}.

\subsection*{Time step-dependent PRNG stream jumping}

PRNG stream ``jumping'' efficiently advances the state of the generator as if a large number of draws had been sampled. On each time step, $k$, we begin by resetting each PRNG stream to its initial state. Then, each PRNG stream is jumped $k$ times. These jumps ensure that each decision-specific stream is in a new unique state that depends only on the initial state and simulation time step.

\subsection*{At most one call to each PRNG per time step}

On each time step, $k$, our approach allows at most one call to each decision-specific PRNG stream. The call may request a large sample size, for example one realization for every agent in the simulation. Limiting the number of calls to at most one on each time step ensures that draws come from a stream that starts from a known state that will be matched between simulations.

\subsection*{Slot-based assignment of random number draws}\label{s:slots}

Each agent in the population is assigned a ``slot'' that is used to index into an array of random numbers drawn from each decision-specific PRNG stream on each time step. Because the slot is used as an index into an array, it must be a non-negative integer.

During initialization of the population of $N$ agents, a column vector of unique identifiers (UIDs) is created and forms the index of the agent matrix. Initial UID values are assigned linearly, $0, 1, \ldots, N-1$. The slot vector for this initial population is simply a copy of the UID vector, so that agent $i$ will receive slot $i$.

On time step $k$, let $\mathcal{S}_d$ be a PRNG stream associated with decision $d$. The stream $\mathcal{S}_d$ has been initialized and jumped according to the rules described above. If any agents are faced with decision $d$ on time step $k$, a vector of $M$ random draws, $\vec{r}$, will be sampled from the stream. The random draws are assigned to individual agents by indexing so that the agent with UID $i$ receives draw $r_{i} = \vec{r}\left[\text{slot}_i\right]$. We set $M = \text{max}(\text{slot}_i)$ for $i$ in the set of agent UIDs faced with decision $d$ on this time step.

We make a distinction between UIDs and slots because new agents may be born into the simulation. While UIDs are assigned sequentially as new agents are added to the population, slots cannot be assigned sequentially because two simulations for which we aspire to achieve CRN may have differing numbers of births. Instead, we determine the slot for newborn agents based on a random number generated by one or both of the biological parents.

Specifically, one of the decisions for which we allocate a PRNG stream is, ``What will be the slot for agents born on this time step?'' Using the slot of the selected parent, we sample a new slot for each linked newborn from a discrete uniform distribution with a lower bound equal to the $N$, the initial population size, and an upper bound equal to $\text{int}(qN)$, where $q>1$ is a user-configurable scalar multiplier with a typical value in the range of 2 to 10.

Because slots are drawn from a discrete uniform distribution, there is a chance that two or more agents could receive the same slot. Two agents with the same slot, facing the same decision, on the same time step will receive the same random realization. The chances of such a collision can be reduced to near-zero by increasing $q$. However, increasing the number of available slots comes at the cost of increasingly large draw sizes because the number of random numbers drawn must be large enough to accommodate all requested slots.

When creating newborn agents, all properties beyond the slot, such as birth sex and any other user-configured covariates, are determined using separate decision streams with stochastic realizations indexed by the slot of the newborn. Thus, two agents born on the same time step who happen to receive the same slot will receive identical properties at birth. Some outcomes experienced by these ``twin'' agents will be identical, like the timing of demographic events, but other outcomes like network edge formation and disease acquisition, sequelae, and onward transmission will differ.

\vspace{\baselineskip}
\noindent\fbox{%
    \parbox{0.9\textwidth}{%
\textbf{Example:} Consider a population of $N=10$ agents in which agents 0, 5, and 8 (these numbers refer to UIDs) have assigned slots 0, 5, and 21. These agents are newly infected, and now we seek to determine the prognosis for each from a Weibull distribution. We draw 22 Weibull-distributed random numbers and only use the draws at the 1\textsuperscript{st}, 6\textsuperscript{th}, and 22\textsuperscript{nd} positions, corresponding to the zero-based slots associated with these agents.
}}
\vspace{\baselineskip}

\subsection*{Pairwise random numbers}\label{s:pairwise}

It is often necessary in disease modeling to have a random draw that acts on a pair of individuals, rather than for each agent individually, for example to determine whether one agent infects another. For $N$ agents, there are $N(N-1)/2$ possible pairwise interactions (i.e. edges), although in networks in practice tend to be sparse. Na{\"i}vely, one would sample an independent random number for each edge in the network. However, such an approach is not CRN safe because the addition or loss of a single agent (or interaction) will cause the random numbers for all subsequent pairs to change.

The innovation we make here is to calculate a uniformly distributed random number $u_{ij}$, used for each pair of agents $i$ and $j$, based on random numbers drawn by agents $i$ and $j$. Specifically, let $u_i$ and $u_j$ be random 64-bit unsigned integers sampled for agents $i$ and $j$ using the techniques described above. We then apply a deterministic transformation, $f(u_i, u_j)$ to yield a uniformly distributed random realization $u_{ij} \in [0,1)$.

After exploring several alternatives and checking for bias, see \nameref{a:pairwise}, we settled on the following transformation,
\begin{align}
    u_{ij} = \text{xor}(u_i * u_j, u_i-u_j) / M_{64},
\end{align}
where $u_i, u_j \sim U(0,M_{64})$ are random 64-bit integers and $M_{64}$ is the largest 64-bit integer.

\subsection*{Network edge formation}

Pathogen transmission within \starsim occurs on edges of a multi-layer dynamic transmission network. Nodes in this network represent individual agents and edges represent contacts. Edges are dynamic and therefore may form and dissolve over time. The network is multi-layer in the sense that users can group edges into ``layers'' representing place (e.g.~home, school, work, community), transmission route (e.g.~airborne, sexual, environmental), relationship type (e.g.~marital, casual, commercial), or other factors.

One of the most challenging aspects of achieving CRN in a dynamic transmission model is maintaining coherence in network connections. Many common network algorithms are not ``CRN safe'' in the sense that the presence or absence of even just one additional agent can cause all new connections formed on that time step to differ.

We have identified three network algorithms that maintain common random number coherence despite possible changes in the number of agents (nodes) available for connections due to birth, death, or other reasons. The three methods we describe here differ significantly in their capabilities and performance scaling.

\subsubsection*{Erd{\H o}s-R{\'e}nyi Dynamic Random Network}
In an Erd{\H o}s-R{\'e}nyi graph, each pair of nodes is connected with probability $p$. Ordinarily, $N(N-1)/2$ random numbers would be used in assessing the existence of an undirected edge between each pair of agents. To create a CRN-safe Erd{\H o}s-R{\'e}nyi network, we instead use one of the pairwise random number methods described in \nameref{s:pairwise} (other than Modulo, which is biased). With this approach, edges can have a defined duration or be can recreated on each time step. The loss or addition of agents will not affect other network edges. We note that this approach could be generalized. For example, the probability, $p$, of an edge could depend on agent properties, simulation time, or other factors.

\subsubsection*{Dynamic Disk Graph}
To create a dynamic disk graph, we initially place each agent randomly on a two-dimensional unit square using the techniques described above. Edges are created between agents that are separated by a distance of $r$ or less, where $r$ is a scalar radius to be determined by the user.

Such a network can be made dynamic by moving the agents on each time step. Here, we have explored two simple approaches. The first is a random walk in which each agent samples a new position from a normal distribution centered at the current position, again using the techniques above.
\begin{equation}
    \begin{bmatrix}x_{t+dt}\\ y_{t+dt} \end{bmatrix} \sim \mathcal{N}\left( \mu=\begin{bmatrix}x_t\\ y_t\end{bmatrix}, \Sigma_t \right),
\end{equation}
for any 2D covariance matrix $\Sigma_t$.

The second approach ascribes a constant velocity, $v$, and orientation, $\theta$, to each agent and calculates new positions as a forward step of length $v * dt$ in direction $\theta$,
\begin{equation}
    \begin{bmatrix}x_{t+dt}\\ y_{t+dt} \end{bmatrix} = \begin{bmatrix}x_t + v \cos(\theta_t) * dt\\ y_t + v \sin(\theta_t) * dt\end{bmatrix},
\end{equation}
where $dt$ is the time step of the model.

With each approach, agents wandering outside of the unit square are reflected back in. With the constant-velocity method, the orientation is updated so that agents ``bounce'' off walls. Other motion updates are possible, each creating different dynamic disk networks that will not be altered by the addition, loss, or other agent-network participation changes.

We note that this gerenal concept could be generalized in many ways, while retaining the CRN-safe property. For example, the user could change away from a two-dimensional square or use a different distance function.

\subsubsection*{Topological Embedding}

While the previous approaches create simple random networks that are safe for use with CRN, they lack the capability to create detailed assortative edges, a limitation the ``topological embedding'' approach seeks to overcome.  Our approach begins by embedding agents seeking an additional connection in a $d$-dimensional normed vector space. Denote by $\vec{x}_i \in \mathbb{R}^d$ the position of agent $i$ within this space. After embedding, we form a distance matrix $D$ with entry $i$, $j$ computed as the distance between respective agents,
\begin{align}
    D_{ij} = \Vert \vec{x}_j - \vec{x}_i \Vert.
\end{align}
Pairs are assigned by solving the linear sum assignment problem~\cite{crouse2016implementing} using $D$ as the cost matrix. As \starsim is written in Python, we are using the \texttt{linear\_sum\_assignment} function from the \texttt{scipy.optimize} library~\cite{2020SciPy-NMeth}.

The process of embedding each agent may be deterministic. For example, each agent may embed at a position corresponding to their current age, sex, and/or other property like geolocation. Alternatively, a user may employ a stochastic embedding, for example each agent could embed at a position determined by a random draw. When the embedding is stochastic, a purpose-specific pseudo random stream is used in combination with slotting, as described above, to ensure the resulting draws are consistent between realizations.

The linear sum assignment forms a maximal pairing that minimizes the sum of the costs. Here, a maximal pairing ensures that the cardinality of the match is as high as possible; no agents that could be matched are left unmatched. The cost minimization means that nearby pairs of agents are more likely to be matched than distant agents. These properties create an ideal situation for common random number alignment between two simulations as changes, like the addition or removal of agents seeking connections on any given time step, will create a perturbation that is not global, but rather local with respect to the embedding.

Finally, note that bipartite networks can be generated using a distance matrix with rows representing agents of one type (e.g.~women) and columns representing agents of the other type (e.g.~men). We use this approach in an example below to simulate a heterosexual HIV transmission network.

The downside of this approach is performance as the linear sum agreement algorithm is $O(N^3)$ in $N$, the number of agents seeking a new contact.

\subsubsection*{Static network}\label{s:static_net}
Any static network will be safe for use with common random numbers. These networks are static in the sense that edges between nodes do not change over time, and can therefore be created in advance of running the simulation. While agents can be removed from the simulation simply by removing adjacent edges, newborn agents will not be connected to any other agents and therefore would not acquire new infections.

\subsubsection*{Complete network}
Another CRN-safe option is the complete graph, with edges added and removed as agents enter and leave the simulation. Again, no random numbers are used in producing this network.

\subsection*{Pathogen transmission}
\label{s:pathogen_transmission}

A second significant challenge in achieving CRN in epidemiological models comes at the stage of pathogen transmission. Na{\"i}vely, a network consisting of $e$ edges would use $2e$ random draws to determine pathogen incidence, one for each possible direction of each edge. However, this approach is clearly not CRN safe as the loss of any one edge would shift the random draw realizations for all subsequent edges.

We have identified several solutions to overcome this challenge. We describe an acquisition-based approach in \nameref{a:acquisition}. The primary approach we have implemented takes advantage of the pair-specific random numbers, as described in \nameref{s:pairwise}. This method works just like the na{\"i}ve approach, but substitutes a pair-specific random realization instead of a random draw from a centralized generator.

\section*{Results}\label{s:results}

To identify use cases within computational epidemiology for which common random numbers provide a meaningful advantage over the traditional centralized approach, we present results from three examples. First, maternal postpartum hemorrhage prevention in a model with births and deaths, but no transmission. Second, vaccination in a susceptible-infected-recovered (SIR) transmission model with a closed population and static network. Finally, voluntary medical male circumcision (VMMC) impact on human immunodeficiency virus (HIV), demonstrating the combination of an open population and dynamic transmission networks.

All results come from \starsim \cite{starsim} (v1.0), an open-source general-purpose health and disease modeling framework.  The methods to achieve CRN, as described in \nameref{s:methods}, have been designed into the \starsim framework. Simultaneously, the framework features the ability to disable CRN, reverting back to a single centralized random number generator for comparison purposes. We present results comparing the following two approaches to generating random numbers.
\begin{itemize}
    \item \textbf{Centralized:} All random numbers used during the simulation come from a single (centralized) random number generator, as is typical in modern agent-based simulation modeling in epidemiology. The stream is NumPy's default Mersenne Twister. We use the same random number seeds for each scenario to reduce variance.
    \item \textbf{CRN:} Random numbers for each decision, by each agent, at each time step use the techniques presented in this paper to achieve common random number alignment. The slot scale parameter, $q$, is set to its default value of $5$ for the PPH example and $10$ for the SIR and VMMC examples. Transmission is achieved using the XOR method of pairwise pseudo-random numbers, as described in Table~\ref{t:combine_rands}.
\end{itemize}

\subsection*{Maternal \& Child Health: Prevention of postpartum hemorrhage}

In Sub-Saharan Africa, the maternal mortality ratio is estimated to be around 500 deaths for every 100,000 live births~\cite{world2023trends}. Postpartum hemorrhage (PPH), defined as losing at least half a liter of blood within 24 hours of delivery, is a leading cause accounting for about 25\% of maternal deaths in this region~\cite{say_global_2014}. A recent clinical trial has demonstrated that a package of interventions can reduce a composite measure of severe outcomes from PPH by 60\%~\cite{gallos_ioannis_randomized_2023}.

Further, when a mother dies, her newborn baby has a significantly reduced chance of surviving the first 42 days~\cite{nguyen_risk_2019}. Estimates of infant mortality vary, but sources indicate that roughly half of infants without a mother will die within this period~\cite{finlay_effects_2015}.

To explore the potential benefits of averting PPH, we apply our approach to common random numbers on a synthetic population resembling sub-Saharan Africa. These results demonstrate the ability of the CRN-based approach to simulate vital dynamics, births and deaths, as made possible by the ``slots'' described in \nameref{s:slots}.

Each simulation begins in 2015 and ends in 2030, corresponding to the end-year of the Sustainable Development Goals. The initial population age structure, age- and year-specific fertility rates, and age-, year-, and sex-specific mortality rates are based on data from UN World Population Prospects~\cite{wpp}.

For demonstration purposes, we suppose a 60\% effective PPH-prevention intervention package was delivered at 10\% or 90\% coverage starting in 2020, 5 years after the beginning of the simulation. We assume the baseline rate of maternal mortality due to PPH is 1 per 1,000 live births and that infants who lose their mother to PPH experience a one-time 50\% chance of death that acts in addition to the baseline mortality rate. Mothers saved by the PPH-prevention package have a knock-on effect of increasing the survival of their newborn.

Each simulation contains 100,000 synthetic agents and we use $250$ replicates by sweeping the random number seed from 0 to 249. The simulation time step is set to one year. We compare centralized and CRN approaches to random number generation.

\begin{figure}[!ht]
\centering
\includegraphics[width=0.9\textwidth]{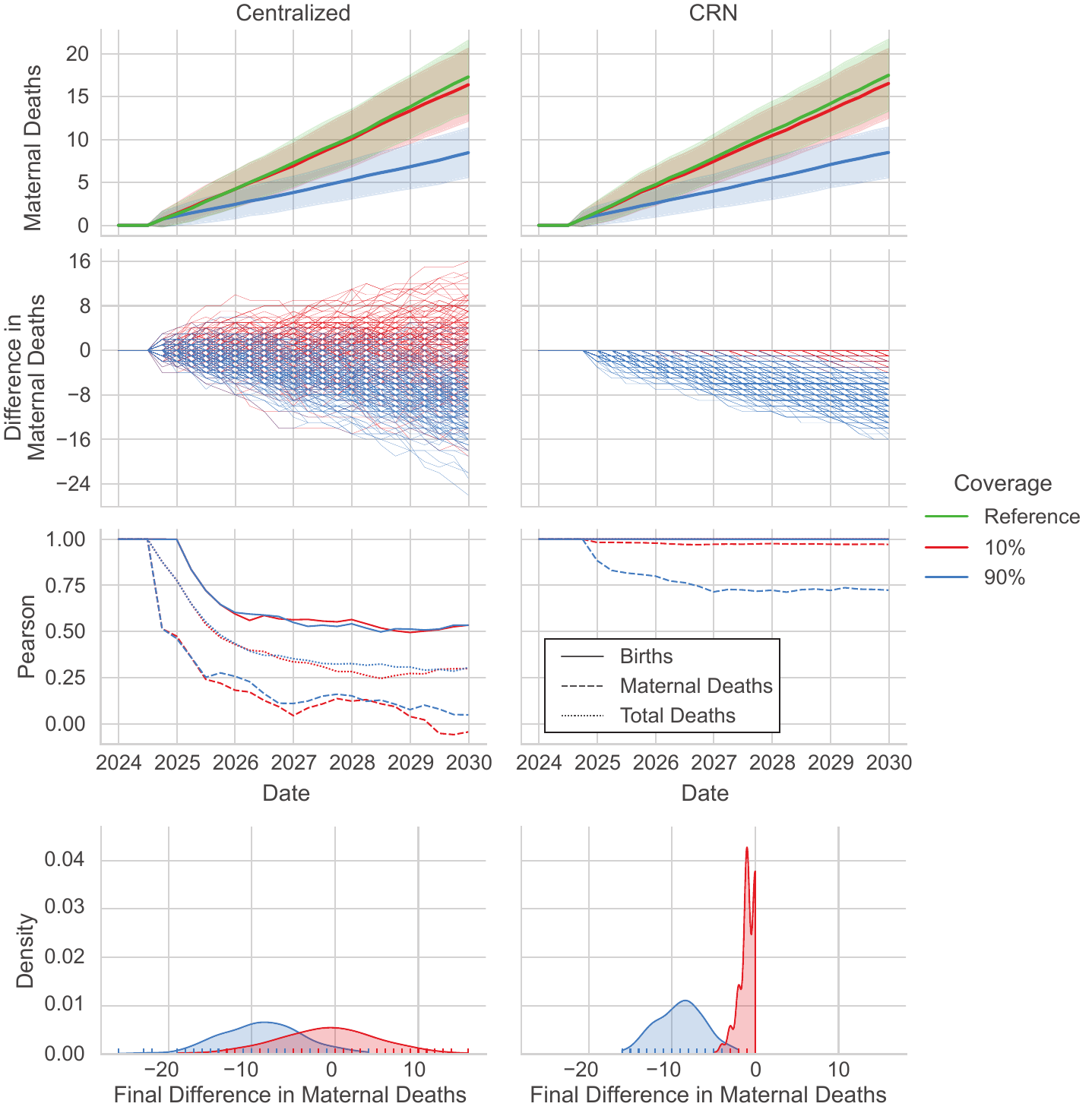}
\caption{Results from PPH simulations comparing centralized (left) and common random number (right) approaches. The top row shows the absolute number of maternal deaths occurring in this synthetic population with the shaded regions corresponding to plus or minus one standard deviation. The second row shows differences between each of the two coverage levels and the reference scenario, paired by random number seed. The third row shows the time-evolution of the Pearson correlation coefficient for three output channels, as indicated by linetype. Finally, the bottom row shows the distribution of differences in maternal deaths at the the final time. Colors indicate coverage of the PPH prevention intervention ranging from 0\% (Reference, green) to 10\% (red) and 90\% (blue).}
\label{f:pph_ts}
\end{figure}

Results displayed in the top row of Fig~\ref{f:pph_ts} show time series trends of cumulative maternal deaths for intervention coverage levels of 10\% and 90\% in addition to the reference, which does not contain any PPH-prevention. The results for the two approaches to random number generation are indistinguishable, suggesting that pseudo random numbers generated by both centralized and CRN approaches are indeed random.

The second row of Fig~\ref{f:pph_ts} shows the differences between the indicated PPH-prevention coverage level and the reference for each of the 250 replicates. In computing these differences over time, we have paired up common random number seeds so that seed $s$ of each coverage level is compared against seed $s$ of the reference. We observe that between-simulation differences are significantly less variable using the CRN approach as compared to the traditional centralized approach. The CRN simulations show a clear decrease in deaths, with larger magnitude for the higher coverage level. Differences using the centralized approach are much more variable, and the trend at 10\% coverage is challenging to discern.

In addition to a significant reduction in variance, the CRN approach demonstrates another advantage. Because differences are realized mechanistically at the individual level instead of in aggregate the population level, a purely beneficial intervention like this PPH-prevention package always results in fewer maternal deaths and more live births, as illunstrated in the bottom row of the results fiture. The same cannot be observed for the centralized approach because real differences are masked by stochastic noise.

The variance of the difference between simulation configurations can be reduced by increasing the covariance term in Eq~\ref{eq:var}. The third row of Fig~\ref{f:pph_ts} shows how the Pearson correlation coefficient (PCC), a measure of covariance, varies as a function of time for several output channels; higher values indicate greater correlation. The CRN approach yields correlation that is significantly higher than the centralized approach at all time points. The drop in PCC for the centralized approach begins early in the simulation, following the first random draw difference between baseline and counterfactual scenarios.

Comparing 10\% and 90\% coverage levels, Pearson correlation with the reference scenario is higher in the 10\% coverage scenario than the 90\% coverage scenario with CRN. In contrast, the centralized approach appears to be insensitive to coverage. Viewing postpartum hemorrhage as a relatively rare event (1 per 1,000 live births), 10\% coverage of the PPH prevention package with 60\% efficacy affects relatively few women and their children, and therefore simulation results for this low-coverage scenario should closely resemble the reference scenario. In other words, the mechanistic signal is small and thus correlation between the 10\% and reference coverage scenarios should be high. In contrast, the 90\% coverage level affects many more women and their babies, and thus correlation between baseline and counterfactual simulations should be lower, as demonstrated in the CRN results.

The CRN approach shows lower correlation for maternal deaths than births, especially at 90\% coverage. This too makes sense in light of the fact that the intervention directly averts maternal deaths whereas any change in births are an indirect consequence of women receiving the PPH prevention package to survive to a subsequent pregnancy. Over longer periods of time, newborns that survived because their mother received the PPH intervention could as well contribute new births into the population.

Finally, we can consider the standard error (SE) and the potential number of simulations saved by CRN in achieving a given SE. When considering maternal deaths at the final time, we find that the CRN approach yields 6.2- and 1.75-fold reductions in standard error for 10\% and 90\% coverage levels, respectively. Due to the square-root relationship between standard error and the number of simulations, this reduction amounts to 38- and 3-fold savings in the number of simulations that would need to be run to achieve a specified level of standard error. These savings are much more substantial when considering the number of births that occur. Here we find 14,000- and 1,800-fold reductions in the number of simulations for a given standard error, again for 10\% and 90\% coverage levels.

Additional correlation results for this PPH example can be found in \nameref{a:PPHcor}.

\subsection*{Infectious disease transmission}

While simulations using CRN will always have less stochastic pseudo-random number noise compared to the traditional centralized approach, the actual benefits of CRN may not be meaningful in situations where in-simulation mixing is large and/or the real signal is small compared to the between-replicate variation. To investigate the limits of the advantages of CRN, we present results on a susceptible-infectious-recovered (SIR) infection process evolving on a static population with static network connections, and consider the impact of introducing a vaccine.

Simulation results presented in this section use SIR dynamics with an exponentially-distributed duration of infection with a mean of 30 days. The infection fatality ratio is set at 5\%. We default to a \ba ``power law'' network topology with parameter $m=1$. Static networks are CRN-safe, recall \nameref{s:static_net}. We select a time step of one day and set the default transmissibility parameter, $\beta$, so that there is a 20\% chance of transmission per day between each connected pair of infected an susceptible agents. The size of the simulated population is varied. Simulations start on the first day of 2020 and end 6-months later. The simulated vaccine, which acts to reduce susceptibility to infection acquisition, is distributed on day five at 5\% or 90\% coverage and is assumed to have a constant efficacy of 70\%.

\begin{figure}[!ht]
\centering
\includegraphics[width=\textwidth]{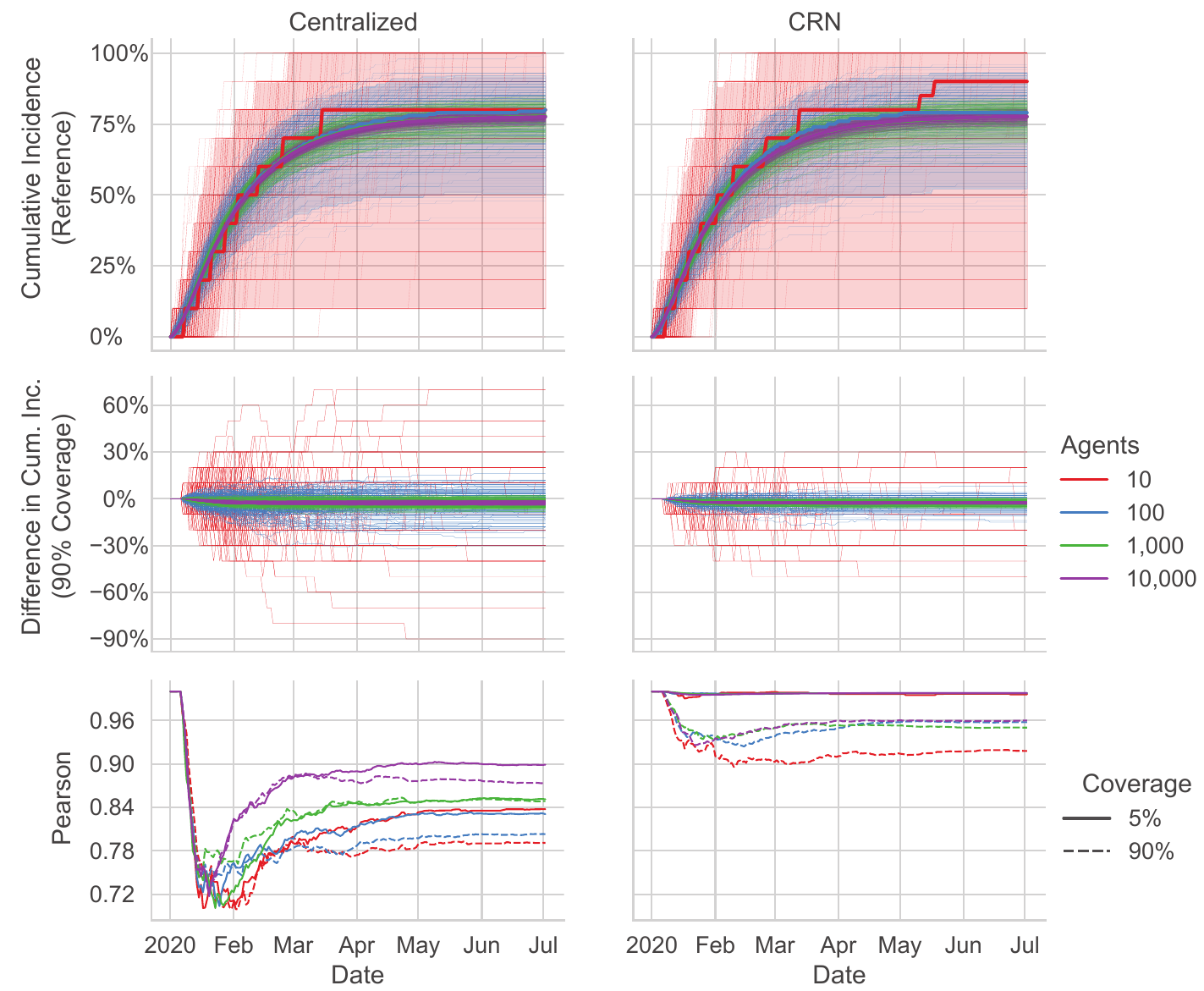}
\caption{Timeseries result for the SIR example for population sizes ranging from 10 (red) to 1,000 (purple) for centralized (left) and CRN (right) approaches to random number generation. The top row shows the cumulative number of infections as a percentage of the population whereas the middle row shows the difference in cumulative incidence between the 90\% coverage level and the no-vaccine reference scenario. The final row shows how the Pearson correlation coefficient against the reference evolves over time for cumulative incidence with 5\% (solid) and 90\% (dashed) vaccine coverage.}
\label{f:sir_n}
\end{figure}

Results comparing cumulative incidence (attack) of centralized and CRN approaches for 10, 100, 1,000, and 10,000 agents for the reference scenario are shown in Fig~\ref{f:sir_n} (top row). As in the PPH example, outputs for both approaches to random number generation yield similar aggregate results. Cumulative incidence of infection is characterized by a high level of quantization and variance for 10 agents, with increasing resolution and decreasing variance as the number of agents is increased, as expected.

The middle row of of Fig~\ref{f:sir_n} show differences between baseline and 90\% coverage, matched by random number seed. The CRN approach appears to have lower variance differences, and differences shrink with increasing number of agents.

To better understand the impact of the population size on the value of CRN, we again turn to the time evolution of the Pearson correlation coefficient, see the bottom row of Fig~\ref{f:sir_n}. These panels clearly show the benefits of the CRN approach. At all time points, the CRN approach results in higher correlation than the traditional centralized approach. Correlation increases with the number of agents. The benefit of CRN over centralized, as quantified by the difference in PCC, narrows as the number of agents increases. But benefits are still apparent at 10,000 agents, especially for the 5\% coverage scenario. As with the PPH example above, correlation is higher at 5\% coverage than 90\% coverage, particularly for CRN, due to the smaller overall perturbation to the system and ability of the CRN approach to avoid loss of correlation due to stochastic noise.

The use of common random numbers results in variance reduction, as can be quantified by the fold-reduction in the number of simulations that need to be run to achieve a given standard error level. Here we find that at 5\% coverage, the use of CRN saves over 10-fold simulations, with nearly 30x savings realized for the smallest population size. The reduction in the number of simulations is smaller for the 90\% coverage level due to the larger signal produced by the intervention. Here we find a modest savings of about 20\%.

Additional results generated by varying the topology of the SIR network are presented in \nameref{a:SIR_topology}.

\subsection*{HIV prevention via voluntary medical male circumcision}

Previous results have explored a dynamic population without transmission (PPH) and a static population with transmission (SIR). We now present results leveraging all aspects of our CRN solution through a simulation of Human Immunodeficiency Virus (HIV) that involves births, natural- and disease-cause deaths, and a dynamic heterosexual disease transmission network.

A recent study conducted the HIV Modelling Consortium evaluated the cost effectiveness of a 5-year continuation of voluntary medical male circumcision (VMMC) as compared to discontinuation of the service over 5-, 20-, and 50-year horizons~\cite{bansi2023cost}. VMMC is estimated to have approximately 60\% efficacy in reducing HIV acquisition in men, but funding for VMMC programming is varied.
 
Motivated by this real-world scenario analysis, we implemented a simple VMMC scenario analysis based on a simple HIV module complete with an age-stratified heterosexual transmission network, mother-to-child transmission, age/sex/year-specific demographics, and temporal scale-up of antiretroviral treatment (ART) and VMMC. HIV prevalence, ART, and VMMC trends were roughly calibrated reflect the epidemiological context in sub-Saharan Africa. Simulated circumcision typically occurs around time of sexual debut and coverage increases to 45\% linearly from 2007 to 2020. The baseline scenario discontinues VMMC in year 2020 whereas the intervention scenario continues VMMC services through 2025. Results come from simulations with an initial population of $10,000$ agents spanning from 1980 to 2070 with a one-month time step and $500$ replicates.

\begin{figure}[!b]
\centering
\includegraphics[width=\textwidth]{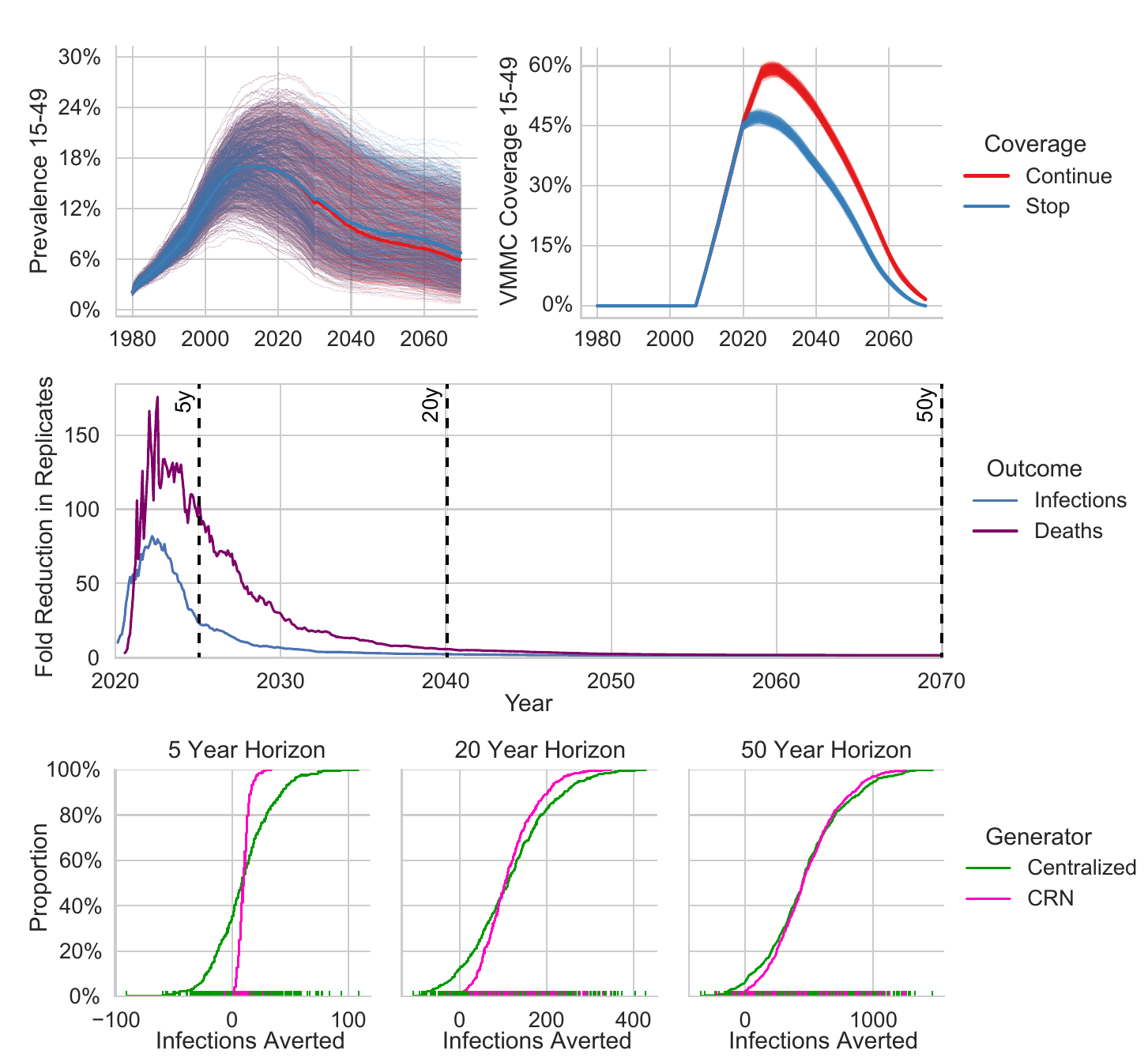}
\caption{Results from the voluntary medical male circumcision example. The top row shows HIV prevalence (left) and VMMC coverage in men (right) in the 15-49 age group. Link color distinguishes scenario between continuing the VMMC program (red) or discontinuing in 2020 (blue). The middle panel shows the fold reduction in the number of simulation replicates that would be required to achieve a given standard error at each point in time during the simulation. Infections averted and deaths averted are displayed in blue and purple, respectively. Dashed vertical lines mark 5, 20, and 50 year horizons. Finally, the bottom row shows cumulative distributions of infections averted 5-, 20-, and 50-years after the beginning of the intervention. Results are presented for the centralized (green) and CRN (pink) generators. }
\label{f:hiv_ts}
\end{figure}

Model outputs showing HIV prevalence, coverage of VMMC, fold reduction in the number of simulations that would be required to achieve a given standard error, and the distribution of infections averted at 5, 20, and 50-year time horizons are presented in Fig~\ref{f:hiv_ts}. As with previous examples, absolute outputs like HIV prevalence and the coverage of VMMC do not differ visually between centralized and CRN approaches, to the top panels display results from the CRN simulations.


The benefit of the CRN approach is most significant over short time horizons. We find that the fold reduction in the number of simulation replicates needed to achieve a given standard error peaks approximately 3-years after the beginning of the intervention. The peak values are approximately 75- and 125-fold reductions in the number of simulations required. Benefits of using CRN fall from these peak levels over time as the impact of the VMMC intervention eventually affects most agents in the system. At 5, 20, and 50-year horizons, the fold reductions are 24, 2.3, and 1.4 for infections averted and 95, 5.7, and 1.6 for deaths averted.

As a final note, we observe in the bottom row of panels in Fig~\ref{f:hiv_ts} that a significant number of replicates generated using the centralized generator have negative infections averted. This result would suggest that continuation of the VMMC intervention is somehow resulting in more infections that would have occurred in the scenario where VMMC is discontinued. Of course, this result is spurious due to random number noise. Indeed, results from the CRN generator show very few replicates with negative infections averted. Unlike the PPH example above in which agents did not mix, here it is possible for the VMMC continuation scenario to result in more infections, even with CRN, due to chains of events that can occur mechanistically as a result of the intervention.

\section*{Discussion}\label{s:discussion}

We present a new methodology for agent-based disease modeling that achieves common random numbers (CRN) to enable precise individual-level comparison between simulations, a critical step towards eliminating unwanted noise when evaluating intervention impact and parameter sensitivity. With our CRN-based approach, two simulations on the same population are comparable at the individual level, and all differences are due to the mechanistic action of the difference. Results show that our approach achieves CRN and always outperforms the traditional centralized approach to random number generation.

The postpartum hemorrhage prevention example demonstrated dramatic increases in signal to noise in an open, but non-interacting, population. We observed up to a four order of magnitude reduction in the number of simulations required to achieve a given standard error as compared to the centralized approach, a huge savings. Results are also easier to communicate as this purely-beneficial intervention cannot possibly result in increased deaths when using the CRN approach.

Transmission examples with SIR dynamics on a static population confirmed that the CRN-based approach is never worse, and revealed that the greatest gains come when the ``signal'' (e.g.~the effect size of the intervention) is small compared to between-run ``noise'' (variance). Gains were more significant for smaller population sizes, but persisted even in a well-mixed simulation of 10,000 agents, a situation where careful use of random numbers might not be expected to yield benefits. When exploring different network topologies with SIR dynamics, we were surprised to see larger gains in Pearson correlation with the faster-mixing \ba topology compared to three other network structures. This power-law network produces high-variance outputs as a result of a the influence of a relatively small number of high-degree nodes. Individual-level alignment between simulations, as enabled by CRN, ensures better alignment of infection reaching high-degree nodes, and therefore more significant variance reduction.

For decades, agent-based models have suffered a signal-to-noise problem. It has been challenging to quantify the impact of interventions reaching select populations and small parameter changes as used in sensitivity analysis. These challenges may have led cost-effective interventions being overlooked. Further, modeling results confused stakeholders with counter-intuitive results showing purely beneficial interventions resulting in worse outcomes for some simulations. Negative and near-zero impact results also complicate cost effectiveness calculations because the incremental impact appears in the denominator.

Our approach is the first to achieve common-random number alignment in agent-based health and disease modeling. Our approach fully eliminates stochastic noise due to misaligned random number realizations. What remains is purely mechanistic signal that can be audited in the sense that every difference in model outputs can be traced back to a physical change in process or parameter value.

Our approach has several limitations.  First, model re-engineering may be required to retrofit an existing agent-based model with the methods described in \nameref{s:methods}. Second, the resulting model code could be more challenging to use and modify. In traditional modeling, a user could call the system ``rand'' function, which accesses a centralized generator; here additional care must be taken by users when evaluating stochastic decisions.

While random number generation typically takes a small percentage of overall simulation time, our slot-based approach to random numbers draws many more random numbers than are actually used. While this seems wasteful and does reduce model performance, random number generation is not a significant performance bottleneck in our experience. Instead, we have found the embedding network has $O(N^3)$ scaling that dramatically affects performance with large $N$. Users seeking more performance or larger population sizes could use the disk or \er algorithms we described. The performance of these algorithms is assessed in \nameref{a:perf}.

Another possible limitation may come from how ``slots'' are assigned to agents. Slots are used to map random number realizations to agents, and there is no guarantee that slots are unique. Agents sharing a slot will receive the same random realizations, possibly leading to undesirable correlations. Our approach to choosing slots allows users to trade off the probability of repeated slots with the number of random numbers that are generated for each decision. Users of these methods should conduct sensitivity and validity analysis to balance this trade off.

Not all agent-based modeling applications benefit equally from our approach. Interventions with large effect sizes or large changes to input parameters generate a large signal for which the benefits of careful random number alignment are less meaningful in practice. Users should weigh the benefits of CRN against performance and complexity considerations.

Finally, while we conducted a variety of simulation experiments, our results explore a relatively small corner of the space of all agent-based modeling applications. We expect our approach to outperform the traditional centralized approach in all applications, but acknowledge our results are limited in this regard.

\section*{Supporting information}

\paragraph*{Software and analysis code availability}
Methods described in this article have been implemented in \starsim, which is available as open source code on GitHub~\cite{starsim}. Results were generated using v1.0.1 of the framework. Analysis code is available GitHub repository available online at \url{https://github.com/starsimhub/crn_paper}.

\paragraph*{S1 Appendix}\label{a:pairwise}
{\bf Pairwise random numbers}

We explored several transformation functions designed to create a single uniform random number, $u_{ij}$ from random numbers produced by each agent in a pair, $u_i$ and $u_j$. The approaches we considered are summarized in Table~\ref{t:combine_rands}. Here, $M_{64}$ and $M_{32}$ are the maximum 64- and 32-bit unsigned integer values.
\begin{table}[hb!]
\caption{Comparison of functions to create pairwise random numbers}
\centering
\begin{tabular}{|c|c|c|c|}
    \hline
    \textbf{Method} & \textbf{Inputs} & \textbf{Function} ($f$) \\
    \hline
    Modulo & $u_i, u_j \sim U(0,1)$ & $\text{mod}(u_i + u_j, 1)$\\
    \hline
    Middle Square & $u_i, u_j \sim U(0,M_{64})$ & $\left( u_i * u_j >> 32 \right) / M_{32}$\\
    \hline
    Bitwise XOR & $u_i, u_j \sim U(0,M_{64})$ & $\text{xor}(u_i * u_j, u_i-u_j) / M_{64}$\\
     \hline
\end{tabular}
\label{t:combine_rands}
\end{table}

The Modulo method computes the modulus of the sum of two uniformly distributed random numbers with $1$, the result of which is uniformly distributed. This simple function was a natural starting point. The Middle Square method is based on John von Neumann’s middle-square random number generator~\cite{neumann1951various}, which takes the middle bits after squaring a ``seed'' number. Here, instead of squaring a single number, we take the the middle $32$ bits from the product of $u_i$ and $u_j$, and normalize by the maximum possible unsigned $32$-bit value. Finally, the Bitwise XOR method computes a bitwise exclusive-or between the product and the difference of $u_i$ and $u_j$ before normalizing the result.

\begin{figure}[!b]
\centering
\includegraphics[width=0.9\textwidth]{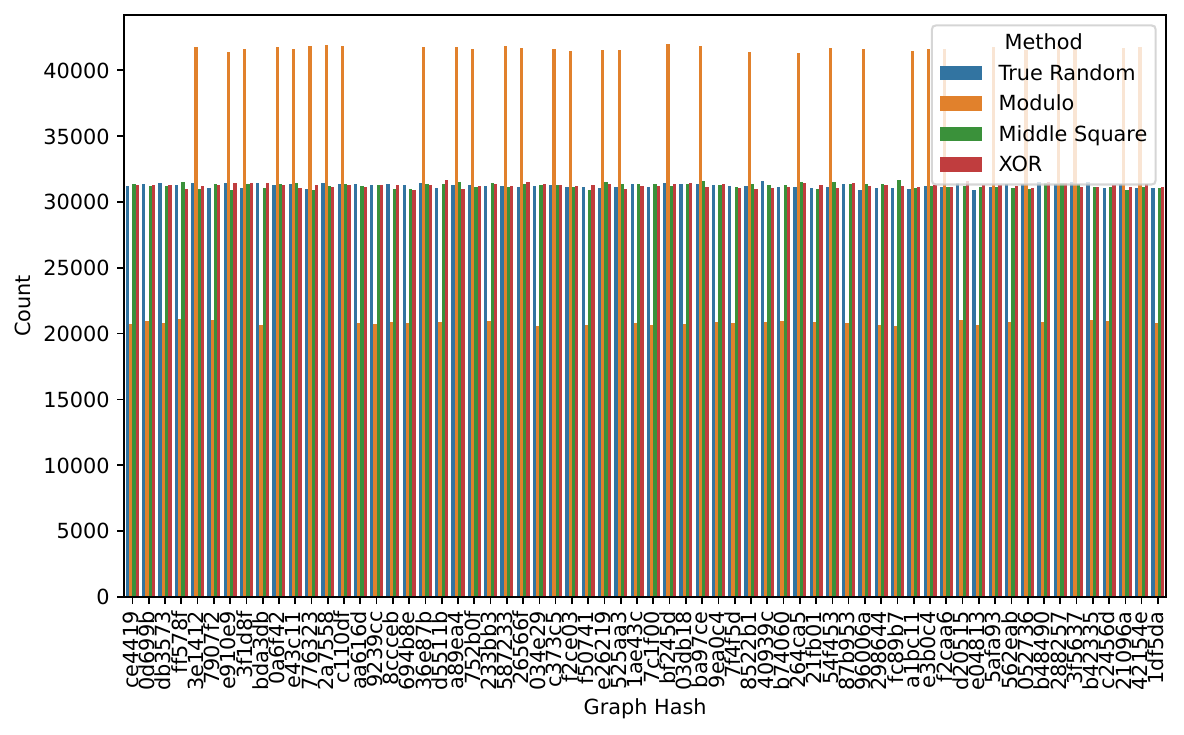}
\caption{Frequencies of random graphs on $N=4$ nodes created using each pairwise approach to random number generation against the ``True Random'' pseudo-random reference. Each hash code on the X-axis corresponds to one unique possible graph. Bar heights show the number of times each graph occurred in two-million replicates. Bar color indexes the method of random number generation used in creating edges.}
\label{f:pairwise_cor}
\end{figure}

For each algorithm, we created 2-million random graphs on $N=4$ and again for $N=6$ nodes. The probability of each edge was set to 50\%. We compared the frequency of resulting graphs against reference frequencies generated using an independent pseudo-random number per edge, which is not CRN safe, from NumPy's default Mersenne Twister implementation. Each resulting random graph was hashed for ease of comparison. Results are presented in Fig~\ref{f:pairwise_cor}.

Already from this bar plot it is clear that the Modulo approach creates a bias. The transmission tree with hash {\tt e3b0c4}, corresponding to no transmissions, is over-represented compared to results generated using non-CRN-safe centralized pseudo-random number generator.

We next employ a statistical test to detect if any of the methods produce biased results. Specifically, we apply a Chi-Squared test of independence to the contingency table between the True Random result and each RNG-safe approach. Results presented in Table~\ref{t:pairwise_cor} show that the Modulo approach can be rejected, but results from the other approaches cannot be rejected as different from True Random.
\begin{table}[h]
    \caption{Results from a Chi-Squared test of independence testing the hypothesis that each method produces transmission trees with a frequency that is not different from (pseudo) True Random.}
    \centering
    \begin{tabular}{|c|c|c|}
    \hline
       \textbf{Method}  & \textbf{p-value} ($\mathbf{N=4}$) & \textbf{p-value} ($\mathbf{N=6}$)\\
    \hline
        Modulo&	0.0 & 0.0\\
    \hline
        Middle Square&	\textcolor{black}{0.78} & \textcolor{black}{0.22}\\
    \hline
        XOR&	\textcolor{black}{0.62} & \textcolor{black}{0.65}\\
    \hline
    \end{tabular}
    \label{t:pairwise_cor}
\end{table}

\paragraph*{S2 Appendix}\label{a:acquisition}
{\bf Acquisition-based disease transmission}

Here, we describe an alternate approach to CRN-safe disease transmission. In this ``acquisition-based'' approach, each node first computes the probability of acquiring infection from any neighbor as,
\begin{equation}
    p_i = 1 - \prod_{j\in\mathcal{N}_i} (1-p_{ij}),
\end{equation}
where $p_{i}$ is the probability of node $i$ acquiring infection from any neighbors, $\mathcal{N}_i$, and $p_{ij}$ is the probability of transmission on the edge between agents $i$ and $j$. With these probabilities in hand, we then use a single random number from agent $i$, applying the techniques described above, to determine if transmission occurs.

The source of each infection can be determined by inverse cumulative transform sampling, again using the techniques described above. While this method is effective and technically sound, we have found it to be slow for some applications, depending on the density of the network.

\paragraph*{S3 Appendix}\label{a:perf}
{\bf Network performance characterization}

A main performance bottleneck in implementing a common random number safe model is the transmission network. In this article, we have proposed three network algorithms that maintain random number alignment between simulations. These three networks vary in their flexibility and performance.

While the ``Embedding'' network allows for user-specified assortative mixing, the use of the linear sum assignment algorithm is a potential performance bottleneck. In comparison, the disk and \er implementations should scale better with increasing number of agents.

To find out, we timed the ``update'' step of each network algorithm for 9 logarithmically spaced population sizes ranging from 10 to 32,000. For each population size, we computed the average time per update for 5 sequential updates, and repeated the experiment with 3 different random number seeds. The experiment was conducted with CRN enabled, but results were not different when using random numbers from a single centralized stream.

\begin{figure}[t!]
\centering
\includegraphics[width=\textwidth]{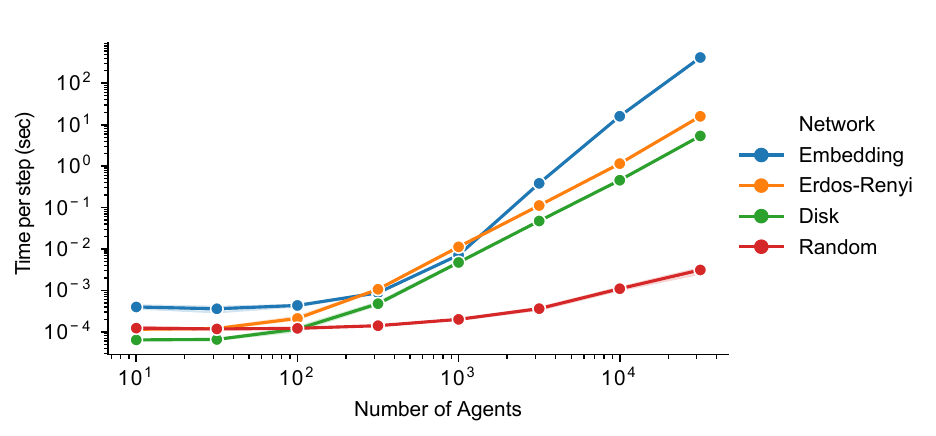}
\caption{Performance scaling of CRN-safe network generation algorithms. We plot the time to generate a network, measured in seconds, as a function of the number of agents from 10 to 32,000. Line color indicates the network generation algorithm. Of the algorithms compared, the Random network is the only one that is not CRN safe. Note the log10 scale on both X- and Y-axes.}
\label{f:net_perf}
\end{figure}

Results presented in Fig~\ref{f:net_perf} illustrate that the Disk and \er algorithms have similar performance and scaling characteristics. Both are an order of magnitude faster than the Embedding network by 10,000 agents. For comparison, we have included performance scaling results from the Random network algorithm, which is not CRN safe. This algorithm is highly performant, primarily relying on an array shuffle operation to create random pairings. Shuffle-based approaches are not safe for use with common random numbers because the addition or removal of even just one agent would cause all subsequent pairings to differ.

Finally, please note that in many applications, edges in the network representing contacts persist for several consecutive time steps, representing lasting relationships, and therefore it is typical for much fewer than the full population $N$ to be seeking additional contacts on any one timestep. Also, users can configure the model so that only some agents are eligible for new network connections on each step. But here, for performance evaluation, we test the networks by ensuring that all agents are eligible for edges on each and every network update, in part by discarding any edges created on previous updates. Thus, this test represents a worst-case scenario as all agents are seeking new edges on every update.

Results were computed on an M1 Macbook Pro. Absolute times will vary with computing hardware, but the relative values and scaling trends will be consistent.

\paragraph*{S4 Appendix}\label{a:PPHcor}
{\bf Additional PPH correlation results}

\begin{figure}[!t]
\centering
\includegraphics[width=\textwidth]{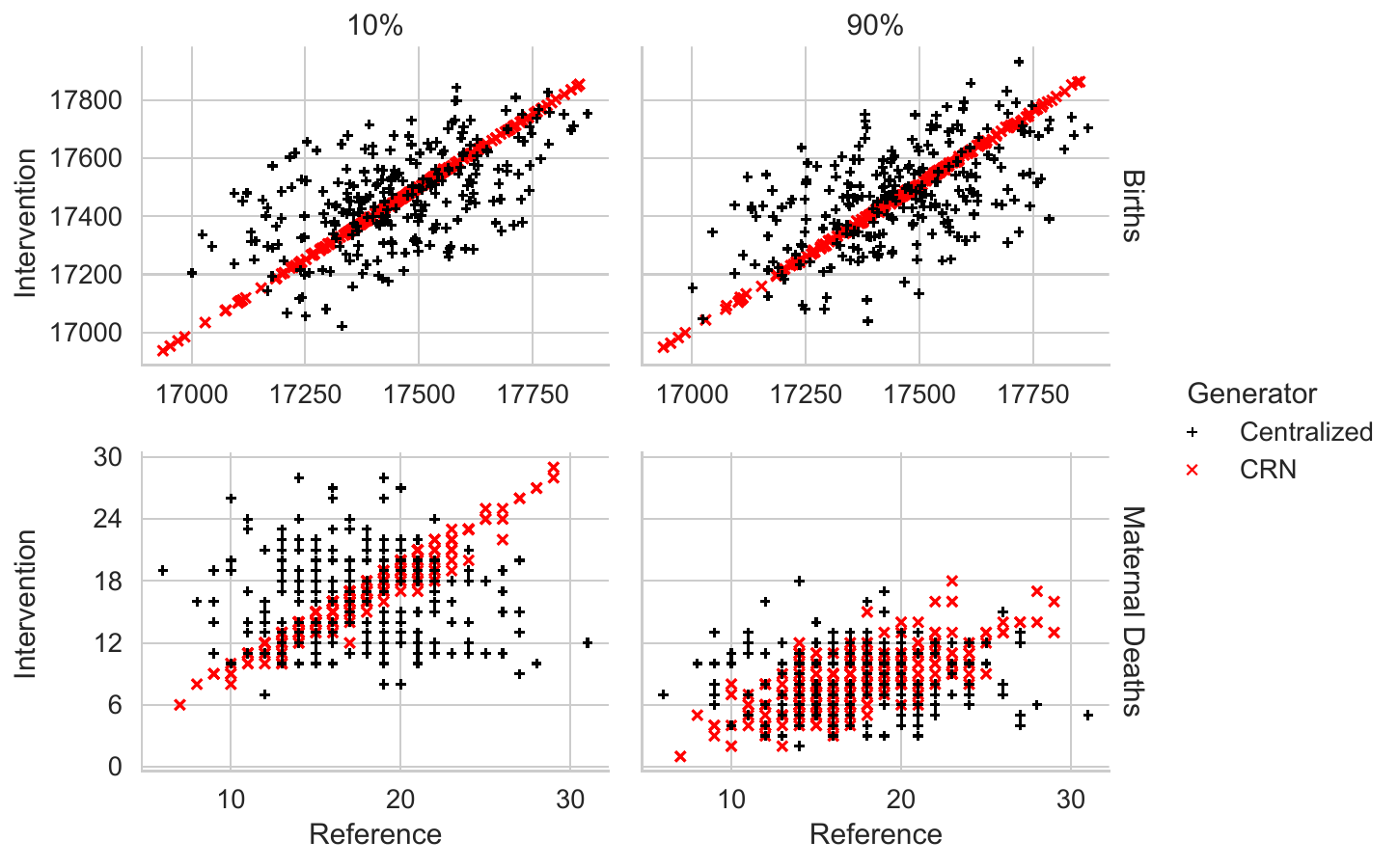}
\caption{Scatter plot comparing cumulative counts of births (top row) and maternal deaths (bottom row) for 10\% (left) and 90\% (right) coverage levels. Individual points represent seed-matched pairs of simulations, the reference result on X and the intervention result on Y. Dot color and shape reference the random generator used.}
\label{f:pph_cor_slice}
\end{figure}

To visualize the enhanced correlation due to reduced random number noise, Fig~\ref{f:pph_cor_slice} shows additional results from the PPH example. In this figure, each dot represents one pair of simulation results at the final time. The value on the X-axis is the number of births (top) and maternal deaths (bottom) in the reference scenario, which does not include any PPH prevention, and the value on the Y-axis is the corresponding number in an intervention simulation generated using the same random number seed.

All results demonstrate a clear correlation, thanks to shared seeds used with the centralized approach and common random numbers for the CRN approach.. Simulations that happen to result in high (low) values without PPH prevention also have high (low) values with PPH prevention. Results generated using the CRN approach (red, ``x'' markers) demonstrate higher correlation than those generated with the centralized approach (black, ``+'' markers) due to removal of unwanted random number noise.

Correlation is higher with 10\% coverage of the PPH intervention (left) because results are more similar to the reference scenario than with 90\% coverage (right). Finally, results show higher correlation for births than maternal deaths due to the fact that births are quickly corrupted by random number noise when using the centralized approach.

\paragraph*{S5 Appendix}\label{a:SIR_topology}
{\bf Varying the SIR network topology}

Understanding that network topology affects mixing, we hypothesize networks inducing slower mixing will favor our approach based on common random numbers. Here, we test this hypothesis using SIR disease dynamics on four static network topologies: \ba with $m=1$, \er with $p=0.004$, Watts-Strogatz with parameters $k=4$ and $p=0.2$, and a 2D planar grid using $1,000$ agents. Simulations run for two years beginning in 2020.

To ensure comparability of results across differing network topologies, we have calibrated the transmissibility parameter, $\beta$, for each network to achieve a final attack of 60\% (600 agents) in the reference scenario at the final time.

\begin{figure}[!t]
\centering
\includegraphics[width=\textwidth]{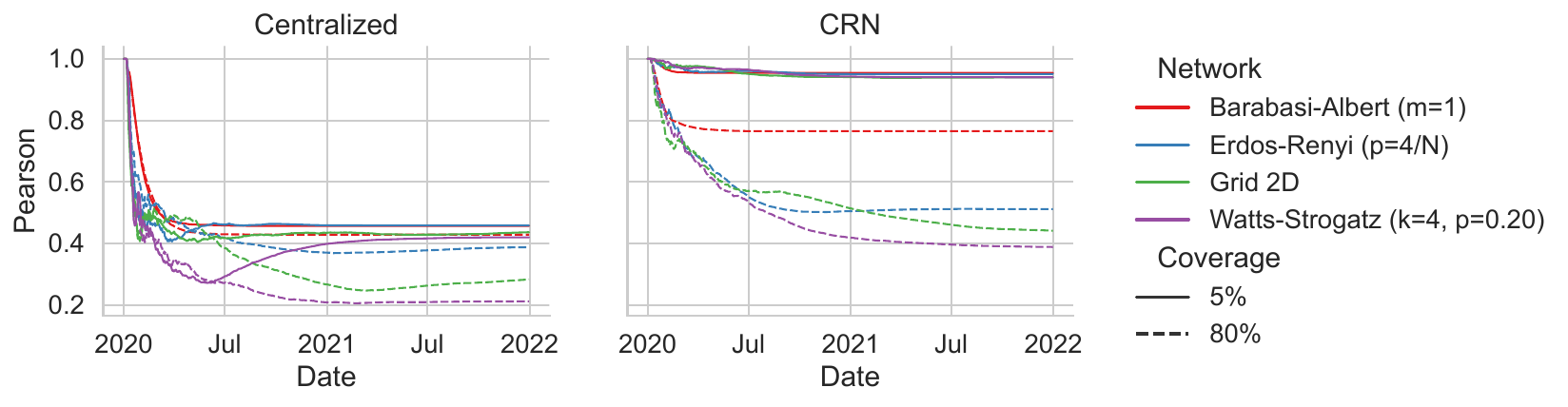}
\caption{Pearson correlation of the number of recovered agents between reference and vaccine scenarios for centralized (left) and CRN (right) approaches. Each line color represents one network topology, as indicated in the legend, and the line style indicates coverage levels of 5\% (solid) and 80\% (dashed).}
\label{f:sir_net_cor}
\end{figure}

The Pearson correlation coefficient is higher for CRN than centralized at all time points for all network topologies considered, see Fig~\ref{f:sir_net_cor}. The benefit of CRN over the traditional centralized approach to agent-based disease modeling, as quantified by the difference in PCC, is largest for the \ba network. Smaller values are observed for the Watts-Strogatz and Grid 2D networks, with the \er topology in the middle.

This result is counter to our hypothesis because the largest benefit is observed in the topology with the fastest mixing. The structure of the \ba ``power law'' network results in high variance, so there is more opportunity early in the spread for common random numbers to reduce variance by eliminating unwanted noise from misaligned random realizations.

Consistent with previous findings, correlation with the reference scenario is higher when the system is less perturbed, as is the case with 5\% compared to 90\% vaccine coverage.

\section*{Acknowledgments}
This advance would not have been possible without contributions from many to the \starsim modeling framework, in which our methods have been implemented and tested. We would also like to acknowledge and thank Edward Wenger and Philip Welkhoff for their support and thoughtful feedback, and Jen Schripsema for helpful edits.


%
%
%


\begin{thebibliography}{10}

\bibitem{botev_variance_2017}
Botev Z, Ridder A.
\newblock Variance reduction.
\newblock Wiley statsRef: Statistics reference online. 2017; p. 1--6.

\bibitem{kahn_methods_1953}
Kahn H, Marshall AW.
\newblock Methods of {Reducing} {Sample} {Size} in {Monte} {Carlo} {Computations}.
\newblock Journal of the Operations Research Society of America. 1953;1(5):263--278.
\newblock doi:{10.1287/opre.1.5.263}.

\bibitem{kleijnen1974statistical}
Kleijnen JP.
\newblock Statistical techniques in simulation. 1974;.

\bibitem{heikes_using_1976}
Heikes RG, Montgomery DC, Rardin RL.
\newblock Using common random numbers in simulation experiments — an approach to statistical analysis.
\newblock SIMULATION. 1976;27(3):81--85.
\newblock doi:{10.1177/003754977602700301}.

\bibitem{conway_tactical_1963}
Conway RW.
\newblock Some {Tactical} {Problems} in {Digital} {Simulation}.
\newblock Management Science. 1963;10(1):47--61.
\newblock doi:{10.1287/mnsc.10.1.47}.

\bibitem{stout_keeping_2008}
Stout NK, Goldie SJ.
\newblock Keeping the noise down: common random numbers for disease simulation modeling.
\newblock Health Care Management Science. 2008;11(4):399--406.
\newblock doi:{10.1007/s10729-008-9067-6}.

\bibitem{murphy_using_2013}
Murphy DR, Klein RW, Smolen LJ, Klein TM, Roberts SD.
\newblock Using {Common} {Random} {Numbers} in {Health} {Care} {Cost}-{Effectiveness} {Simulation} {Modeling}.
\newblock Health Services Research. 2013;48(4):1508--1525.
\newblock doi:{10.1111/1475-6773.12044}.

\bibitem{cornejo_creating_2014}
Cornejo D, Mayorga ME, Lich KH. Creating common patients and evaluating indiviual results: Issues in indivual simulation for health policy analysis; 2014.

\bibitem{flaxman_untangling_2017}
Flaxman AD, Deason AW, Dolgert AJ, Mumford JE, Sorensen RJ, Eldrenkamp E, et~al.. Untangling uncertainty with common random numbers: a simulation study; 2017.

\bibitem{kaminsky_perfect_2019}
Kaminsky J, Keegan LT, Metcalf CJE, Lessler J.
\newblock Perfect counterfactuals for epidemic simulations.
\newblock Philosophical Transactions of the Royal Society B: Biological Sciences. 2019;374(1776):20180279.
\newblock doi:{10.1098/rstb.2018.0279}.

\bibitem{kerr2021covasim}
Kerr CC, Stuart RM, Mistry D, Abeysuriya RG, Rosenfeld K, Hart GR, et~al.
\newblock Covasim: an agent-based model of COVID-19 dynamics and interventions.
\newblock PLOS Computational Biology. 2021;17(7):e1009149.

\bibitem{fpsim}
O’Brien ML, Valente A, Kerr CC, Proctor JL, Noori N, Root ED, et~al.
\newblock FPsim: An agent-based model of family planning.
\newblock npj Women's Health. 2023;1(1):1.

\bibitem{hpvsim}
Stuart RM, Cohen JA, Kerr CC, Mathur P, Abeysuriya RG, Zimmermann M, et~al.
\newblock HPVsim: An agent-based model of HPV transmission and cervical cancer.
\newblock PLOS Computational Biology. 2024;.

\bibitem{starsim}
{Cliff Kerr, Robyn Stuart, Romesh Abeysuriya, Paula Sanz-Leon, Jamie Cohen, and Daniel Klein}. Starsim;.
\newblock Available from: \url{https://github.com/starsimhub/starsim}.

\bibitem{pcg}
O’neill ME.
\newblock PCG: A family of simple fast space-efficient statistically good algorithms for random number generation.
\newblock ACM Transactions on Mathematical Software. 2014;.

\bibitem{crouse2016implementing}
Crouse DF.
\newblock On implementing 2D rectangular assignment algorithms.
\newblock IEEE Transactions on Aerospace and Electronic Systems. 2016;52(4):1679--1696.

\bibitem{2020SciPy-NMeth}
Virtanen P, Gommers R, Oliphant TE, Haberland M, Reddy T, Cournapeau D, et~al.
\newblock {{SciPy} 1.0: Fundamental Algorithms for Scientific Computing in Python}.
\newblock Nature Methods. 2020;17:261--272.
\newblock doi:{10.1038/s41592-019-0686-2}.

\bibitem{world2023trends}
Organization WH, et~al.
\newblock Trends in maternal mortality 2000 to 2020: estimates by WHO, UNICEF, UNFPA, World Bank Group and UNDESA/Population Division: executive summary. 2023;.

\bibitem{say_global_2014}
Say L, Chou D, Gemmill A, Tuncalp O, Moller AB, Daniels J, et~al.
\newblock Global causes of maternal death: a {WHO} systematic analysis.
\newblock The Lancet Global Health. 2014;2(6):e323--e333.
\newblock doi:{10.1016/S2214-109X(14)70227-X}.

\bibitem{gallos_ioannis_randomized_2023}
{Gallos Ioannis}, {Devall Adam}, {Martin James}, {Middleton Lee}, {Beeson Leanne}, {Galadanci Hadiza}, et~al.
\newblock Randomized {Trial} of {Early} {Detection} and {Treatment} of {Postpartum} {Hemorrhage}.
\newblock New England Journal of Medicine. 2023;389(1):11--21.
\newblock doi:{10.1056/NEJMoa2303966}.

\bibitem{nguyen_risk_2019}
Nguyen DTN, Hughes S, Egger S, LaMontagne DS, Simms K, Castle PE, et~al.
\newblock Risk of childhood mortality associated with death of a mother in low-and-middle-income countries: a systematic review and meta-analysis.
\newblock BMC Public Health. 2019;19(1):1281.
\newblock doi:{10.1186/s12889-019-7316-x}.

\bibitem{finlay_effects_2015}
Finlay JE, Moucheraud C, Goshev S, Levira F, Mrema S, Canning D, et~al.
\newblock The {Effects} of {Maternal} {Mortality} on {Infant} and {Child} {Survival} in {Rural} {Tanzania}: {A} {Cohort} {Study}.
\newblock Maternal and Child Health Journal. 2015;19(11):2393--2402.
\newblock doi:{10.1007/s10995-015-1758-2}.

\bibitem{wpp}
of~Economic UND, Social~Affairs PD.
\newblock World Population Prospects 2024; 2024.
\newblock Available from: \url{https://population.un.org/wpp/}.

\bibitem{bansi2023cost}
Bansi-Matharu L, Mudimu E, Martin-Hughes R, Hamilton M, Johnson L, Ten~Brink D, et~al.
\newblock Cost-effectiveness of voluntary medical male circumcision for HIV prevention across sub-Saharan Africa: results from five independent models.
\newblock The Lancet Global Health. 2023;11(2):e244--e255.

\bibitem{neumann1951various}
Neumann V.
\newblock Various techniques used in connection with random digits.
\newblock Notes by GE Forsythe. 1951; p. 36--38.

\end{thebibliography}

\end{document}